%% file: main.tex
\documentclass{JHEP3}
\usepackage{graphicx}
\usepackage{amsmath}\usepackage{amssymb}\usepackage{bbm}
\makeatletter
\newdimen\rh@wd
\newdimen\rh@hta
\newdimen\rh@htb

\def\rh@measure#1{\setbox\rh@box=\hbox{$#1$}\rh@wd=\wd\rh@box
\rh@hta=\ht\rh@box}

\def\widecheck#1{\rh@measure{#1}%
   \setbox\rh@box=\hbox{$\widehat{\vrule height \rh@hta width\z@
\kern\rh@wd}$}%
   \rh@htb=\ht\rh@box \advance\rh@htb\rh@hta \advance\rh@htb\p@
   \ooalign{$\vrule height \ht\rh@box width\z@ #1$\cr
            \raise\rh@htb\hbox{\scalebox{1}[-1]{\box\rh@box}}\cr}}
\makeatother
\makeatletter
\def\widebar{$\m@th\mathord=\mkern-7mu
  \cleaders\hbox{$\!\mathord=\!$}\hfill
  \mkern-7mu\mathord=$}
\makeatother

\title{Symmetries of  
Self-Dual Yang-Mills Equations Dimensionally Reduced from (2,2) Space-time} 
\author{Paul Mansfield  \\
Department of Mathematical Sciences, University of Durham\\\newbox\rhbox
South Road, Durham, DH1 3LE, U.K.\\ 
E-mail: \email{p.r.w.mansfield@durham.ac.uk}
}

\author{Adam Wardlow  \\
Department of Mathematical Sciences, University of Durham\\\newbox\rhbox
South Road, Durham, DH1 3LE, U.K.\\ 
E-mail: \email{a.b.wardlow@durham.ac.uk}
}

\abstract{We construct infinite-dimensional symmetries of the two dimensional equation which results from the dimensional reduction of the self-duality condition in 
 $(2,2)$ signature space-time. These are symmetries of the dimensionally reduced Chalmers-Siegel action and so hold off-shell.}
\preprint{DCPT-10/51}
\begin{document}
\numberwithin{equation}{section}
\input{introduction}

\input{Hitchin}

\input{light_cone_equations}

\input{2d-symmetries}

\input{summary}
\acknowledgments
A.W. wishes to thank the STFC for a studentship. P.M.  thanks STFC for support under the rolling grant ST/G000433/1. We also thank Niall MacKay for pointing out the connection to earlier work on conserved currents in sigma-models.

\appendix

\bibliography{references} 
\bibliographystyle{jhep}
\end{document}

%% file: introduction.tex
\section{Introduction}

We shall construct infinite dimensional symmetries of the 
action
\begin{equation}
\nonumber \text{Tr}\int d^2 x \overline{\Psi}\left (\left(\partial_u^2+\partial_z^2\right)\Psi-\overline{\Psi}\left[\partial_z\Psi , \partial_u\Psi\right]\right)
\end{equation}
with Euler-Lagrange equation
\begin{equation}
 \nonumber (\partial_u^2+\partial_z^2)\Psi-[\partial_z\Psi,\partial_u\Psi]=0
\end{equation}
resulting from the dimensional reduction of the self duality condition in $(2,2)$ space. The self-dual Yang-Mills action written in four dimensions has physically relevant solutions and in particular in Euclidean space the solutions are referred to as instantons. Starting with the Euclidean equations in $\mathbb{R}^4$ they may be dimensionally reduced by demanding that solutions are invariant under translations of two of the coordinates, \cite{Hitchin-86}. The solutions of the dimensionally reduced equations, called Hitchin's equations, have the property that they may be defined over a Riemann surface using analytic maps and are conformally invariant, and they have found applications in the field of integrability amongst others; see for example \cite{Kneipp:2008dc,Bochicchio:2003kw,Ivanova:1991tu} and \cite{Jardim}. We review the procedure taken by Hitchin in \cite{Hitchin-86}.

In the paper, \cite{Mansfield:2009ra}, it is shown how to construct infinite symmetries of the Chalmers Siegel action \cite{Chalmers:1996rq} describing the self-duality equations in $(1,3)$ space using (complex) light-cone coordinates,  $x_o=t-x_3$,  $x_{\bar{o}}=t+x_3$, $z=x_1+ix_2$, $\bar{z}=x_1-ix_2$ for $t$, $x_1$, $x_2$ and $x_3$ real. (See also \cite{Dolan2,PhysRevD.25.1086,Preitschopf,Popov:1998pc,Ivanova:2000zt} for previous discussions on symmetries of Yang-Mills theories) This work was based on the derivation of a Lagrangian formalism to the MHV rules, \cite{Parke:1986gb,boels-2007-648,Cachazo:911222,Cachazo:2004kj}, in the papers \cite{ettle_and_morris,Pauls_paper} and was extended to the $N=4$ supersymmetric action in the papers \cite{huang} and \cite{Wardlow:2009hk}. The procedure involves defining a canonical transformation that maps the Chalmers Siegel action to a free theory. By writing symmetries of the free theory and using the transformation and its inverse the authors construct symmetries of the self-dual action. In particular, it was mentioned in \cite{Mansfield:2009ra} that by making $x_2$ imaginary, the light-cone system became a set of real coordinates and the arguments in that paper extended to $(2,2)$ space-time where $A_z$ and $A_{\bar{z}}$ are not related by complex conjugation. By writing the Euclidean Cartesian coordinates $(x_1,x_2,x_3,x_4)$ in terms of two complex coordinates, $z=x_1+ix_2$ and $u=x_3+ix_4$ we write the Hitchin equations by assuming that the fields do not depend on the imaginary parts of the new coordinates, $x_2$ and $x_4$ where we assume an anti-hermitian representation of the Lie algebra valued fields with $A_{\bar{u}}^*=-A_u=$ and $A_{\bar{z}}^*=-A_z$. In the case where $x_2$ and $x_4$ are pure imaginary then we arrive at the same dimensionally reduced equations but now the fields are no longer related by complex conjugation and on the plane we may impose the gauge $A_{\bar{u}}=0$. Since the symmetry $\delta A_z$, (written down in the introduction to \cite{Mansfield:2009ra}) with real momenta is a symmetry of the 4d action in $(2,2)$ space (using the light-cone gauge), where fields are not related by conjugation, it is necessarily a symmetry of the equation of motion. By dimensionally reducing the expression, we write down an expression for the symmetry of the 2d equation of motion.

%% file: Hitchin.tex
\section{The Hitchin system}
The Hitchin equations result from the dimensional reduction of the self-duality equations in Euclidean space, \cite{Hitchin-86}. We shall review their derivation using real space-time coordinates $(x_1,x_2,x_3,x_4)$, before introducing the complex co-ordinates that allow immediate generalisation to (2,2) signature. So the metric is $ds^2=dx_1^2+dx_2^2+dx_3^2+dx_4^2$.
Hitchin considers the Lie-algebra valued curvature two form
\begin{equation}
\nonumber  F(A)=dA+A^2\sum_{\mu < \nu}F_{\mu\nu}dx^{\mu}\wedge dx^{\nu}
\end{equation}
where $A$ is the connection over the G-bundle
\begin{equation}
 \nonumber A=A_{1}dx^{1}+A_{2}dx^{2}+A_{3}dx^3+A_{4}dx^{4},
\end{equation}
and we choose an anti-hermitian representation of the Lie algebra generators, $T^a$, so that if $A=A_a\,T^a$, then $A_a^{*}=A_a$ and $A^{*}=-A$.
When written out explicitly the self-duality condition,
\begin{equation}
 \nonumber F_{\mu\nu}=\frac{1}{2}\sqrt{g}\varepsilon_{\mu\nu\lambda\rho}F^{\lambda\rho},
\end{equation}
becomes the set
\begin{equation}
 F_{12}=F_{34},\quad
 F_{13}=F_{42},\quad
F_{14}=F_{23}.
\end{equation}
Hitchin, \cite{Hitchin-86}, then assumes that the functions $A_{\mu}$ are independent of two of the coordinates, $x_2$ and $x_4$ say, (note, for later convenience,  this is a different choice form that of \cite{Hitchin-86}). So that the connection becomes
\begin{equation}
 \nonumber A=A_{1}dx^{1}+A_{3}dx^{3}.
\end{equation} 
Re-naming the fields $A_{2}$ and $A_{4}$ as
\begin{equation}
 \nonumber A_{2}=\varphi,\ \ \ \ \ \ \ \ \ \ A_{4}=\psi
\end{equation}
which are reminiscent of Higgs fields, \cite{Hitchin-86}, puts the equations of motion into the form
\begin{align}
 \nonumber F_{12}&=\left[D_1,\varphi\right]=\left[D_3,\psi\right]=F_{34}\\
 \nonumber F_{13}&=\left[D_1,D_3\right]=\left[\psi,\varphi\right]=F_{42}\\
\nonumber F_{14}&=\left[D_1,\psi\right]=\left[\varphi,D_3\right]=F_{23}.
\end{align}

%% file: light_cone_equations.tex
\section{The Self-Duality Equations in Complex and Light-Cone Co-ordinates}

We have seen the derivation of the Hitchin equations in Cartesian coordinates. The approach in the papers \cite{Pauls_paper}, \cite{ettle_and_morris} and \cite{Mansfield:2009ra} is to write the self-dual action and its symmetries in $(1,3)$ light-cone momentum space coordinates, $\check{p}=p_t-p_3$, $\hat{p}=p_t+p_3$, $\tilde{p}=p_1+ip_2$ and $\bar{p}=p_1-ip_2$ where $\tilde{p}$ and $\bar{p}$ are related by complex conjugation. However, it was discussed in \cite{Mansfield:2009ra} that we can make $p_2$ pure imaginary thereby making all $(\check{p},\hat{p},\tilde{p},\bar{p})$ real. Then the arguments written down in \cite{Mansfield:2009ra} extend to $(2,2)$ space. In fact the authors used this to derive their results to begin with.

With that in mind, we write the self-duality equations that are dimensionally reduced from Euclidean space-time in a complex coordinate system. We define complex coordinates $(u,\,\bar{u},\,z,\,\bar{z})$ for Euclidean space by
\begin{align}
\label{eq:euclideanlightconecoords}
\nonumber u&=x_3+ix_4, &   z&=x_1+ix_2\\
\bar{u}&=x_3-ix_4,   & \bar{z}&=x_1-ix_2   
\end{align}
so now 
\begin{equation}
 \nonumber ds^2=dx_1^2+dx_2^2+dx_3^2+dx_4^2=dud\bar{u}+dzd\bar{z}
\end{equation}
The self-duality equations 
\begin{equation}
  \nonumber F_{\mu\nu}=\frac{1}{2}\sqrt{g}\varepsilon_{\mu\nu\lambda\rho}F^{\lambda\rho},
\end{equation}
with $\varepsilon_{u\bar{u}z\bar{z}}=1$, written out in components are now
\begin{equation}
F_{u\bar{u}}=-F_{z\bar{z}},\quad 
F_{uz}=0,\quad 
F_{\bar{u}\bar{z}}=0,
\end{equation}
or more explicitly, 
\begin{align}
\label{eq:selfdualequationsincomplexcoords}
\nonumber \partial_uA_z-\partial_zA_u+[A_u,A_z]&=0\\
\nonumber \partial_{\bar{u}}A_{\bar{z}}-\partial_{\bar{z}}A_{\bar{u}}+[A_{\bar{u}},A_{\bar{z}}]&=0\\
\partial_uA_{\bar{u}}-\partial_{\bar{u}}A_u+[A_u,A_{\bar{u}}]&=-\partial_zA_{\bar{z}}+\partial_{\bar{z}}A_z-[A_z,A_{\bar{z}}]
\end{align}
where the fields are related by conjugation with $A_{\bar{u}}^*=-A_u$ and $A_{\bar{z}}^*=-A_z$ because of the anti-hermitian generators. Following Hitchin's procedure, we assume that the fields are independent of the coordinates $x_2$ and $x_4$. Then we have
\begin{align}
\label{eq:dimredcond}
\nonumber \partial_{x_4}& A_{\mu}=\partial_uA_{\mu}-\partial_{\bar{u}}A_{\mu}=0\\
\partial_{x_2}&A_{\mu}=\partial_{\bar{z}}A_{\mu}-\partial_zA_{\mu}=0
\end{align}
for all the field components, $A_{\mu}$. The Hitchin equations (\ref{eq:selfdualequationsincomplexcoords}) in this coordinate system, are then
\begin{align}
\label{eq:hitchsd1} \partial_uA_z-\partial_zA_u+[A_u,A_z]&=0\\
\label{eq:hitchsd2} \partial_uA_{\bar{z}}-\partial_zA_{\bar{u}}+[A_{\bar{u}},A_{\bar{z}}]&=0\\
\label{eq:hitchsd3}\partial_uA_{\bar{u}}-\partial_u A_u+[A_u,A_{\bar{u}}]&=-\partial_zA_{\bar{z}}+\partial_zA_z-[A_z,A_{\bar{z}}]
\end{align}
and the Higgs fields are,  
\begin{align}
 \nonumber \varphi&=\frac{i}{2}(A_{\bar{z}}-A_z)\\
\nonumber \psi&=\frac{i}{2}(A_{\bar{u}}-A_u)\,.
\end{align}

In the papers \cite{ettle_and_morris,Pauls_paper} and \cite{Mansfield:2009ra} we had a time-like co-ordinate and were able to use a light-cone gauge condition.
This would be analogous to $A_{\bar{u}}=0$ which we are not free to choose in Euclidean space because the components are related by $A_{\bar{u}}^*=-A_{u}$.  So now we consider the problem in $(2,2)$ space by writing $x_2=iy$ and $x_4=it$, with $t$ and $y$ real. Then the coordinates $(u,\,\bar u,\,z,\,\bar z)$ are all real and independent. Furthermore, the fields $A_u$ and $A_{\bar{u}}$ are no longer related by complex conjugation, and neither are $A_z$ and $A_{\bar{z}}$. 

In an appropriate domain we can now make the gauge choice $A_{\bar{u}}=0$ and then by (\ref{eq:hitchsd3}) we can set $A_{\bar{z}}=0$ (without constraining $A_z$). Then using (\ref{eq:hitchsd1}) we can write
\begin{equation}
\label{eq:equationforAu}
A_u=-\frac{\partial_z}{\partial_u} A_z
\end{equation}
and substitute into (\ref{eq:hitchsd2}) to arrive at
\begin{equation}
\label{eq:equationforAz}
 \left(\partial_{u}+\frac{\partial_z^2}{\partial_u}\right)A_z-\left[\frac{\partial_z}{\partial_u} A_z,A_z\right]=0
\end{equation}
or, to make a simplification by defining $A_z=\partial_u \Psi$, we can write
\begin{equation}
\label{eq:2deqninpsi}
 \left(\partial_{u}^2+\partial_z^2\right)\Psi-\left[\partial_z\Psi,\partial_u\Psi\right]=0.
\end{equation}

%% file: 2d-symmetries.tex
\section{Symmetries of the 2D Euclidean Self-Dual Equations on the plane}

\label{tx:2dsymm}
In \cite{Mansfield:2009ra}, infinite dimensional symmetries of the self-dual action in $(1,3)$ space were constructed but by assuming real momenta then the results are valid in $(2,2)$ space. The procedure was extended to the $N=4$ supersymmetric theory in \cite{Wardlow:2009hk}. The Chalmers-Siegel
action that gives rise to the self-duality equation for $A=A_z$ as its Euler-Lagrange equation is
\begin{equation}
\label{eq:action}
S=\frac{4}{g^2}\text{Tr}\int d^4 x \overline{A}\left(\partial_u\partial_{\overline{u}}+\partial_z\partial_{\overline{z}}\right)A+ \frac{4}{g^2}\text{Tr}\int d^4 x\left(\partial_{\overline{z}}\partial_{\overline{u}}^{-1}A\right)\left[A , \partial_{\overline{u}}\overline{A}\right]
\end{equation}
We define the notation, $(p_u^n,p_{\bar{u}}^n,p^n_z,p^n_{\bar{z}})=(\check{n},\hat{n},\tilde{n},\bar{n})$ and $\zeta_p=\bar{p}/ \hat{p}$, and 
\begin{equation}
\nonumber	\int_{1 \cdots n}=\int \frac{d^4 p_1}{(2\pi)^4} \cdots \frac{d^4 p_n}{(2\pi)^4}.
\end{equation}
Writing the action in momentum space gives
\begin{eqnarray}
 \nonumber \text{Tr}\int_1 \left\{\bar{p}_1\tilde{p}_1+\hat{p}_1\check{p}_1\right\}\overline{A}_{\bar{1}} A_1-i \text{Tr}\int_{123} \hat{p}_1\left(\zeta_3-\zeta_2\right)\overline{A}_{\bar{1}}A_{\bar{2}}A_{\bar{3}}(2\pi)^4\delta(p_1+p_2+p_3)
\end{eqnarray}
where we use the shorthand $A_p=A(p)$ and $A_{\bar{p}}=A(-p)$, and we perform the transformation $A\rightarrow A'=A+\varepsilon\delta A$ and $\bar{A}\rightarrow \bar{A}'+\varepsilon\delta\bar{A}$ with
\begin{eqnarray}
	\nonumber\delta A_1=-\varepsilon\sum_{n=2}^{\infty}\sum_{i=2}^{n}\sum_{j=i}^n\int_{2\cdots n}\frac{\hat{1}}{\hat{q}}\Gamma(q^G,i^G,\cdots,j^G)\Gamma(q,j+1,\cdots,n,1\cdots,i-1)\times\\
\nonumber	\times A_{\bar{2}}\cdots A_{\bar{i}^G}\cdots A_{\bar{j}^G}\cdots A_{\bar{n}}\\
\end{eqnarray}
and 
\begin{eqnarray}	
\nonumber \delta \overline{A}_1=-\varepsilon\sum_{n=2}^{\infty}\sum_{k=2}^n\sum_{i=2}^{k-1}\sum_{j=i}^{k-1}\int_{2\cdots n}\frac{\hat{k}^2}{\hat{1}\hat{q}}\Gamma(q^G,i^G,\cdots,j^G)\Gamma(q,j+1,\cdots,n,1,\cdots,i-1)\times\\
\nonumber\times A_{\bar{2}}\cdots A_{\bar{i}^G}\cdots A_{\bar{j}^G}\cdots \overline{A}_{\bar{k}}\cdots A_{\bar{n}}\\
\nonumber+\varepsilon\sum_{n=2}^{\infty}\sum_{k=2}^n\sum_{i=2}^{k}\sum_{j=k}^{n}\int_{2 \cdots n}\frac{\hat{q}}{\hat{1}}\frac{(\hat{k}^{G^{-1}})^2}{(\hat{q}^{G^{-1}})^2}\Gamma(q^{G^{-1}},i^{G^{-1}},\cdots,j^{G^{-1}})\Gamma(q,j+1,\cdots,n,1,\cdots,i-1)\times\\
\nonumber\times A_{\bar{2}}\cdots A_{\bar{i}^{G^{-1}}}\cdots \overline{A}_{\bar{k}^{G^{-1}}}\cdots A_{\bar{j}^{G^{-1}}}\cdots A_{\bar{n}}\\
\nonumber-\varepsilon\sum_{n=2}^{\infty}\sum_{k=2}^n\sum_{i=k+1}^{n}\sum_{j=i}^{n}\int_{2 \cdots n}\frac{\hat{k}^2}{\hat{1}\hat{q}}\Gamma(q^G,i^G,\cdots,j^G)\Gamma(q,j+1,\cdots,n,1,\cdots,i-1)\times\\
\nonumber \times A_{\bar{2}}\cdots \overline{A}_{\bar{k}}\cdots A_{\bar{i}^G}\cdots A_{\bar{j}^G}\cdots A_{\bar{n}}.
\end{eqnarray}
where $\Gamma(12\cdots n)$ was first written in \cite{ettle_and_morris} as
\begin{equation}
\nonumber \Gamma(1,\cdots,n)=-(i)^n\frac{\hat{1}}{(1,2)}\frac{\hat{1}}{(1,2+3)}\cdots\frac{\hat{1}}{(1,2+\cdots (n-1))}.
\end{equation}	
and $p\rightarrow p^G$ is an isometry. The bracket $(i,j)$ is defined as $(i,j)=(\hat{i}\tilde{j}-\tilde{i}\hat{j})$. The invariance of the action under these transformations is proven in \cite{Mansfield:2009ra}. Given that they are symmetries of the action, that is sufficient for us to be able to infer that they are indeed symmetries of the equation of motion which we derived earlier (\ref{eq:equationforAz}). We could have also derived this equation using the action (\ref{eq:action}).

Here, we are concerned with finding symmetries of the dimensionally reduced equation (\ref{eq:2deqninpsi}) where, as before, we have defined $A_z=A=\partial_u\Psi$,
\begin{equation}
 \nonumber \left(\partial_u^2+\partial_z^2\right)\Psi-\left[\partial_z\Psi,\partial_u\Psi\right]=0.
\end{equation}
which in momentum space is
\begin{equation}
\nonumber (\check{1}^2+\tilde{1}^2)\Psi_1-\int d^2p_1 d^2 p_2 (2,3)\Psi_{\bar{2}}\Psi_{\bar{3}}\delta^2(p_1+p_2+p_3).
\end{equation}
where $d^2 p=d\check{p}d\tilde{p}$

By writing $A_{p}=-i\hat{p}\Psi_{p}$ the expression for $\delta A$, (written in the introduction to \cite{Mansfield:2009ra}), becomes 
\begin{eqnarray}
\label{eq:deltapsi4d}
	\nonumber\delta \Psi_1=-\varepsilon\sum_{n=2}^{\infty}\sum_{i=2}^{n}\sum_{j=i}^n\int_{2\cdots n}\frac{\hat{q}}{\hat{1}}D(q^G,i^G,\cdots,j^G)D(q,j+1,\cdots,n,1,\cdots,i-1)\times\\
	\nonumber\times \Psi_{\bar{2}}\cdots \Psi_{\bar{i}^G}\cdots \Psi_{\bar{j}^G}\cdots \Psi_{\bar{n}}\delta^4(\sum_m p_m)\\
\end{eqnarray}
where $D(1,2,\cdots, n)$ is given by
\begin{equation}
\label{eq:hitchD} D(12\cdots n)=-(-1)^n\frac{\hat{1}^{n-3}\hat{2}\hat{3}\cdots\hat{n}}{\left(1,2\right)\left(1,2+3\right)\cdots\left(1,2+3+\cdots+(n-1)\right)}
\end{equation}
in momentum space. The $\Psi(p_1,p_2,p_3,p_4)$ are the Fourier transform of $\Psi(x_1,x_2,x_3,x_4)$,
\begin{equation}
 \nonumber \Psi(p_1,p_2,p_3,p_4)=\int \frac{d^4x}{(2\pi)^4}e^{ip_{\mu}x^{\mu}}\Psi(x_1,x_2,x_3,x_4).
\end{equation}
Since we assume that $\Psi$ depends only on $x_1$ and $x_3$, we have
\begin{align}
 \nonumber \Psi(p_1,p_2,p_3,p_4)&=\int\frac{dx^2}{2\pi}e^{ip_2x^2}\int\frac{dx^4}{2\pi}e^{ip_4 x^4} \int \frac{d^2x}{(2\pi)^2}e^{ip_1x^1+ip_3x^3}\Psi(x_1,x_3)\\
\nonumber &=\delta(p_2)\delta(p_{4})\Psi'(p_1,p_3)\\
\nonumber &=\delta(\hat{p}-\check{p})\delta(\tilde{p}-\bar{p})\Psi'(\check{p},\tilde{p}).
\end{align}
Substitute this into (\ref{eq:deltapsi4d}) and evaluate the integrals over $\hat{p}_i$ and $\bar{p}_i$ for $i=2,\cdots, n$ and we have
\begin{eqnarray}
	\nonumber\delta \Psi_1=-\varepsilon\sum_{n=2}^{\infty}\sum_{i=2}^{n}\sum_{j=i}^n\int_{2\cdots n}\frac{\hat{q}}{\hat{1}}D(q^G,i^G,\cdots,j^G)D(q,j+1,\cdots,n,1,\cdots,i-1)\times\\
	\nonumber \times \Psi'_{\bar{2}}\cdots \Psi'_{\bar{i}^G}\cdots \Psi'_{\bar{j}^G}\cdots \Psi'_{\bar{n}}\delta^2(\sum_m  p_m)\delta(\hat{p_1}-\check{p_1})\delta(\tilde{p_1}-\bar{p_1})
\end{eqnarray}
where now 
\begin{equation}
\nonumber	\int_{1 \cdots n}=\int \frac{d^2 p_1}{(2\pi)^2} \cdots \frac{d^2 p_n}{(2\pi)^2}
\end{equation}
and the kernels, $D$, (\ref{eq:hitchD}) are written in terms of $\check{p}$ and $\tilde{p}$ using $\check{p}=\hat{p}$ and $\tilde{p}=\bar{p}$. Then write the inverse Fourier transform, $\delta\Psi(x_1,x_2,x_3,x_4)$, of $\delta\Psi(p_1)$,
\begin{equation}
 \nonumber \delta\Psi(x)=\int d^4p_1 e^{-ip_{1\mu}x_1^{\mu}}\delta\Psi(p_1)
\end{equation}
and evaluating the integrals over $\hat{p}_1$ and $\bar{p}_1$ the final expression is
\begin{eqnarray}
	\nonumber\delta \Psi'_1=-\varepsilon\sum_{n=2}^{\infty}\sum_{i=2}^{n}\sum_{j=i}^n\int_{2\cdots n}\frac{\hat{q}}{\hat{1}}D(q^G,i^G,\cdots,j^G)D(q,j+1,\cdots,n,1,\cdots,i-1)\times\\
	\nonumber \times \Psi'_{\bar{2}}\cdots \Psi'_{\bar{i}^G}\cdots \Psi'_{\bar{j}^G}\cdots \Psi'_{\bar{n}}\delta^2(\sum_m  p_m)
\end{eqnarray}
and the isometry $x\rightarrow x^G$ is simply a rotation of the plane about angle $t$, viz
\begin{align}
 \nonumber u^{G_t}&=\cos(t) u+\sin(t) z\\
\nonumber z^{G_t}&=-\sin(t)u+\cos(t)z.
\end{align}
Given two transformations, $\delta_{t_1}$ and $\delta_{t_2}$ and the argument in \cite{Mansfield:2009ra}, the commutator of the transformations is
\begin{equation}
\nonumber \left[\delta_{t_1},\delta_{t_2}\right]\Psi=0
\end{equation}
and the infinite set of transformations clearly forms an Abelian algebra. Similarly, by defining $\bar{\Psi}=\partial_u\bar{A}$, we can write the transformation $\delta\bar{\Psi}$ as
\begin{eqnarray}	
\nonumber \delta \overline{\Psi}_1=-\varepsilon\sum_{n=2}^{\infty}\sum_{k=2}^n\sum_{i=2}^{k-1}\sum_{j=i}^{k-1}\int_{2\cdots n}\frac{\hat{q}}{\hat{1}}D(q^G,i^G,\cdots,j^G)D(q,j+1,\cdots,n,1,\cdots,i-1)\times\\
\nonumber\times \Psi_{\bar{2}}\cdots \Psi_{\bar{i}^G}\cdots \Psi_{\bar{j}^G}\cdots \overline{\Psi}_{\bar{k}}\cdots \Psi_{\bar{n}}\\
\nonumber+\varepsilon\sum_{n=2}^{\infty}\sum_{k=2}^n\sum_{i=2}^{k}\sum_{j=k}^{n}\int_{2 \cdots n}\frac{\hat{q}}{\hat{1}}D(q^{G^{-1}},i^{G^{-1}},\cdots,j^{G^{-1}})D(q,j+1,\cdots,n,1,\cdots,i-1)\times\\
\nonumber\times \Psi_{\bar{2}}\cdots \Psi_{\bar{i}^{G^{-1}}}\cdots \overline{\Psi}_{\bar{k}^{G^{-1}}}\cdots \Psi_{\bar{j}^{G^{-1}}}\cdots \Psi_{\bar{n}}\\
\nonumber-\varepsilon\sum_{n=2}^{\infty}\sum_{k=2}^n\sum_{i=k+1}^{n}\sum_{j=i}^{n}\int_{2 \cdots n}\frac{\hat{q}}{\hat{1}}D(q^G,i^G,\cdots,j^G)D(q,j+1,\cdots,n,1,\cdots,i-1)\times\\
\nonumber \times \Psi_{\bar{2}}\cdots \overline{\Psi}_{\bar{k}}\cdots \Psi_{\bar{i}^G}\cdots \Psi_{\bar{j}^G}\cdots \Psi_{\bar{n}}.
\end{eqnarray}

%% file: summary.tex
\section{Summary}
We reviewed the derivation of Hitchin's equations on Euclidean space using complex coordinates. These give an immediate generalisation to $(2,2)$ space-time,
resulting in the two-dimensional equation
\begin{equation}
\nonumber
 \left(\partial_{u}^2+\partial_z^2\right)\Psi-\left[\partial_z\Psi,\partial_u\Psi\right]=0.
\end{equation}
because in this signature we are free to impose a light-cone gauge condition, as in \cite{Mansfield:2009ra}. By
dimensionally reducing the results of that paper we obtained symmetries of this two-dimensional equation $\delta\Psi$ (and $\delta{\overline{\Psi}}$) and, importantly, the action whose Euler-Lagrange equation this is,
\begin{equation}
\nonumber S=\frac{4}{g^2}\text{Tr}\int d^2 x \overline{\Psi}\left (\left(\partial_u^2+\partial_z^2\right)\Psi-\overline{\Psi}\left[\partial_z\Psi , \partial_u\Psi\right]\right)
\end{equation}
where $\bar{\Psi}=\partial_u\bar{A}$.

E. Br\'{e}zin (\textit{et-al}), \cite{brezin} consider two dimensional non-linear sigma models and construct infinite, on-shell, conserved currents of the motion iteratively, subject to the equations of motion being
\begin{align}
 \nonumber \partial_{\mu}A^{\mu}=0, &&  [\partial_{\mu}+A_{\mu},\partial_{\nu}+A_{\nu}]=0,
\end{align}
and the authors give a number of examples of such systems. The system we have studied is an example of such a model\footnote{We thank Niall MacKay for bringing this to our attention.} since, from (\ref{eq:equationforAu}) and $A_z=\partial_u\Psi$, we have $A_u=-\partial_z\Psi$ and 
\begin{align}
 \nonumber \partial_z A_z+\partial_u A_u&=0\\
\nonumber [\partial_z+A_z,\partial_u+A_u]&=0
\end{align}
using (\ref{eq:2deqninpsi}). Our symmetries of the action extend the program of \cite{brezin} and also \cite{Preitschopf,Dolan2,PhysRevD.25.1086} off-shell.

In the introduction to this paper we alluded to the fact that the solutions to the Hitchin equations may be defined over Riemann surfaces using analytic maps. It would be interesting to consider how we might extend our approach to find symmetries of the self-dual equations on such surfaces, for example Riemann spheres or tori.